# Lyapunov Function based adaptive network signal control with deep reinforcement learning

Chaolun Ma[a], Zihao Li[a], Xiubin Bruce Wang [a]*, Yunlong Zhang[a] and Ahmadreza Mahmoudzadeh[a]

[a] *Zachry Department of Civil and Environmental Engineering, Texas A&M University, College Station, TX, USA;*

*corresponding author

Email: bwang@tamu.edu

# Lyapunov Function based adaptive network signal control with deep reinforcement learning

*Abstract* – In traffic signal control, flow-based (optimizing the overall flow) and pressure-based methods (equalizing and alleviating congestion) are commonly used but often considered separately. This study introduces a unified framework using Lyapunov control theory, defining specific Lyapunov functions respectively for these methods. We have found interesting results. For example, the well-recognized back-pressure method is equal to differential queue lengths weighted by intersection lane saturation flows. We further improve it by adding basic traffic flow theory. Rather than ensuring that the control system be stable, the system should be also capable of adaptive to various performance metrics. Building on insights from Lyapunov theory, this study designs a reward function for the Reinforcement Learning (RL)-based network signal control, whose agent is trained with Double Deep Q-Network (DDQN) for effective control over complex traffic networks. The proposed algorithm is compared with several traditional and RL-based methods under pure passenger car flow and heterogenous traffic flow including freight, respectively. The numerical tests demonstrate that the proposed method outperforms the alternative control methods across different traffic scenarios, covering corridor and general network situations each with varying traffic demands, in terms of the average network vehicle waiting time per vehicle.

**INTRODUCTION**

Millions of travelers experience delay at signalized intersections daily, a growing concern in urban traffic management (Schrank et al., 2015). Signal control is crucial to ensuring orderly traffic flow and maximizing intersection throughput or minimizing travel delay (Srinivasan et al., 2006). Traditionally, signal control has been approached through fixed-time, actuated, and adaptive control methods. Fixed-time control relies on predetermined sequences and durations, drawing on historical data without adapting to real-time traffic conditions. Actuated control, conversely, varies green time intervals in response to real-time traffic, allowing for a more responsive timing. Adaptive signal control takes actuated control as a special case but pushes the scope further by leveraging on both current traffic situation and historical record.

The data-rich, sensor-enabled traffic environment has catalyzed the development of adaptive control through means such as Reinforcement Learning (RL). Reinforcement learning (RL) is a highly effective method for discovering near optimal control for nonlinear dynamic and stochastic systems (Buşoniu et al., 2018), but its application is impeded by challenges in the state and the reward function design. Acknowledging the limitations from using only machine learning (ML) for complex scientific and engineering problems, researchers are now exploring the integration of physics models with ML techniques, aiming for well-grounded theoretical approaches (Karniadakis et al., 2021; Radaideh et al., 2021; Willard et al., 2022). Our research here builds on the Lyapunov control, develops into the "max pressure" approach, and proposes an advanced reward function specifically tailored for RL. This proposed reward function synergizes with the traditional traffic control in principle while allowing RL of the multifaceted nature in the context of urban traffic networks. RL can also enhance delay performance under system stability, a task challenging for theoretical analysis. More generally, our investigation has uncovered the Lyapunov function as a unifying platform for several established adaptive signal control methods. This Lyapunov control as a unifying platform can interpret some existing methods as either flow rate maximiser.

This paper integrates the traffic control theory with RL, primarily builds on the Lyapunov function to start with. We claim as one of our contributions is the derivations of Lyapunov functions to the flow-based and pressure-based control. To include RL to pick up application specific network and flow features is another contribution of ours. The test results show robustness and applicability of the derived control, indicating possibly more, additional variations of the Lyapunov control.

**Literature review**

Traffic signal timing came to intense study in the late 1950s with rapidly deteriorating roadway traffic. Urban traffic signal control at the beginning debuted as a network problem, well above the then technical and theoretical capacity. A special case also as a popular practical means, green bandwidth maximization algorithm was among the early methods to generate timing plans to deal with major corridors, equivalently breaking the network problem into one dominated by major traffic corridors while ignoring the minor lines. The bandwidth method tries to maximize the green time intervals along a corridor in which vehicles can progress without stop to its maximum capacity (Ficklin, 1969; Petterman, 1947). Morgan and Little first formulate the bandwidth maximization optimization problem as a mixed integer linear programming (MIP) problem and develop MAXBAND algorithm on arterial and network case (Little, 1966; Little et al., 1981; Morgan & Little, 1964). Gartner et al. (Gartner, 1982; Gartner et al., 2002) develops MULTIBAND, which optimizes all the signal control variables and bandwidth progressions on individual link. The second algorithm to develop timing plans for an arterial street is based on minimizing the delay and number of stops, called flow profile methods. The methods have been applied in commercial software such as SYNCHRO, TRANSYT-7F (Robertson, 1969), and PASSER V. PASSER V uses both algorithms to maximize progression or minimize total delay under both under-saturated and over-saturated traffic conditions when optimized over a set of possible phase sequences (N.A. Chaudhary et al., 2002).

Optimization over traffic situation, most likely together with predicted future traffic variation becomes the essence of adaptive control. Corridor green band maximization mentioned above represents a relatively successful means in practice. However, the literature behind adaptive control is much broader and has been enriched overtime. SCOOT is one of the initial strategies, optimizes traffic signals by using real-time street data to estimate future stops and delays, making minor adjustments to signal timings like splits, offsets, and cycle lengths to minimize traffic disturbances. (Stevanovic et al., 2009). Optimized policies for adaptive control (OPAC) strategy transcends the limitations of fixed-value parameters by continuously optimizing the control system to enhance vehicle throughput at intersections, instead of relying on periodic local controller adjustments (Gartner et al., 2002). There are methods applied to general network control. For example, RHODES claims improvements by integrating upstream vehicle data with stop-line detectors for each of the approaches to calculate loads on links and predict future platoon size and route choices (Mirchandani & Head, 2001). Varaiya introduces the max pressure (MP) concept borrowed directly from hydraulic system control, using the differential queue length between the current and downstream queues to minimize queue differentials and it proves the MP method maintains network stability and is asymptotically optimal (Varaiya, 2013). MP requires minimal data to make it relatively more applicable, but it is not optimal in itself (Wei et al., 2018). Network control is implemented at individual intersections. Treating the network as a collection of isolated intersections appears probably the most primitive and likely also the most practical way of network control. DORAS-Q has emerged as a means of dealing with an isolated intersection with potential to partially consider connections between adjacent intersections in real time [20]. It determines phase changes based on an efficiency metric that considers current queues, short-term arrival predictions, and historical rates for all phases. Its control tends to drive switch to the next most efficient phase. These developments epitomize the ongoing quest to refine traffic signal control through the integration of real-time traffic data, predictive analytics and some (kind of) optimization method.

To briefly summarize, adaptive traffic control typically minimizes queues and delay at intersections through a process of state transition. Stochastic factors or incomplete data make it complicated to accurately describe the state transition. State approximation to guide control for near-optimal objectives remains the dominant way. It may take the form of an approximated mathematical objective function. It may consider a complete network or reduce it to one or more isolated corridors while ignoring subsidiary roads or intersections, in which case the performance measure becomes maximizing the green bandwidth on major arteries. Other general performance measures include minimizing the average network travel time of vehicles and maximizing the total vehicle throughput on the network.

However, approximation as mentioned in the traditional state transition approximation is easily said than done. Traffic fluctuation over time of day or over days of week poses a great challenge to modelers. Also, the network layout and vehicular behavior may inherently imply correlation between upstream and downstream traffic with perturbation from side streets, which is pervasive and hard to capture in often highly simplified models. How to capture the uncertainty and stochasticity remains a challenge hard to address in the scientific fields (C. Wang et al., 2018, 2019; Xiao et al., 2022). This inspires us to resort to reinforcement learning within the general framework of vastly improved computational power and artificial intelligence. Reenforced learning has become a powerfully applied practical tool to automatically learn about correlation and patterns between time series and data across locations. It fits the environment of network traffic control with numerous uncertain factors but likely with numerous repetitive patterns beyond simple human grasp due to the numerous granular details such as vehicle location and speed, number of lanes, speed limit, queue length, headway, vehicle itineraries, etc. Therefore, RL presents a promising venue for enhancing traffic signal control (Abdoos et al., 2014; Abdulhai et al., 2003; El-Tantawy et al., 2013; Chen et al., 2020; Hu et al., 2022; Li et al., 2023; Mo et al., 2022). RL is an advanced machine learning approach where an agent interacts with its environment to perform certain actions, which are then reinforced by rewards to learn the most effective strategies. Leveraging computational advancements, RL can effectively optimize traffic signal control by using the rich data to continually learn and adapt. Yet, the challenge in the RL framework lies in the careful definition of state, action, and the reward function. Traditional environment states in traffic control have such factors as queue length, waiting time, volume, speed, the position of vehicles, delay, phase, and duration (Wei, Zheng, et al., 2019). The criterion for a good design of state and reward is to enable the agent to extract useful information for the optimization. There is a rich literature dedicated to finding the optimal reward function alone (Araghi et al., 2013; Jin & Ma, 2015; Teo et al., 2014; Yau et al., 2017). The common idea to define the reward function is to use a weighted sum of state components such as queue length, and waiting time (Wei et al., 2018), which generally does not reach their minimum through the control process.

Incorporating delay or travel time into a reward function in Reinforcement Learning (RL) presents a notable challenge, despite their apparent alignment with the objective of primary measures. These variables are not always readily available in real time. Moreover, they are susceptible to a range of external factors, including free-flow speed, platoon dispersion, travel behavior, and vehicle types. Let alone a degree of stochasticity in all the above factors. The inherent randomness may end up mapping the states to unrealistic rewards, thus fail the model by preventing it from convergence. In other words, there is a gap in relating the reward with the conventional performance measures in traffic flow theories. The insight here is that the measures incorporated shall be observable after each control action. A pressing need is to align the heuristics including the reward function in RL with the traffic flow theories. Those reward functions include link congestion (e.g. critical road density) (Xu et al., 2020) and difference between upstream and downstream flows (Wei, Chen, et al., 2019). Selection of performance measures shall facilitate the RL optimization in the traffic signal control.

Additionally, most current traffic signal control assumes homogenous passenger car traffic at the intersection (Benekohal & Zhao, 2000; Kong et al., 2022; Zhao & Ioannou, 2016). However, freight traffic are likely to cause much longer delay freight vehicle's unique operational characteristics, which is particularly true along major freight corridors. Representing the volume of mixed traffic is through converting the traffic to the passenger car equivalents (PCE) by using adjustment factors. PCE was first introduced in the US Highway Capacity Manual (HCM) to illustrate the effect of the truck on traffic stream according to headway ratio (Scraggs, 1964). HCM 2010 (Elefteriadou, 2016) assumes a PCE value of 2.0 for heavy vehicles approaching signalized intersections. Recently, freight traffic has been treated as noise or perturbation into state space to examine the robustness of Reinforcement Learning methods on multi-modal traffic signal control (Tan et al., 2020).

This study proposes a RL framework to design a new traffic control policy considering freight traffic in major corridors and network.

The major contributions of this research include:

*Unified Theoretical Framework*: We show that several representative network control may be derived from Lyapunov functions, each defined differently with a difference focus.

*Theoretically Founded RL Method*: the study introduces a reinforcement learning-based method into intersection signal control that is not only data-driven but also organically integrated with traffic flow theories.

*Flexibility in Diverse Traffic Scenarios*: It handles a variety of traffic scenarios, including corridor/network configurations, fluctuating traffic demands, and heterogeneous traffic flows.

The following section describes our developed methodology. We start with different definitions of the Lyapunov function to show it as a unifying platform among several established signal control methods and algorithms. We then proceed to introduce the reinforcement learning (RL) method, and specifically detail the reward in its design to be aligned with the traffic flow theories. In the end, the proposed RL based network control is evaluated through simulations in arterial and grid networks, and its performance is benchmarked against several established methods. The paper concludes with the section of results and findings.

**Methodology**

Lyapunov control refers to the use of a Lyapunov function to adaptively control a dynamical system through a process of state transition. A Lyapunov function, $L$ must be non-negative in $\mathcal{R}$, and meet two key conditions. Firstly, it should act as a non-negative scalar measure of the system's multidimensional state vectors, reflecting system performance. Secondly, the function's time derivative along the system's trajectories should consistently be negative, indicating that the function's value increases as the system shifts towards a less desired state. A drift is defined as the change of the Lyapunov function from time $t$ to $t + \Delta t$. By steering the drift towards a negative value, the function gradually moves closer to zero, suggesting the system approaches stability. The choice of a particular Lyapunov function is at the user's discretion, designed to best represent the state of the system. Different forms of the function can be applied to the same system, giving rise to varying control policies and performances. In this study, we examine two different definitions of the Lyapunov function, one being the sum of squared queues and the other being the sheer sum of queue lengths over the network intersections. We find that each of them leads to a different control policy with a different performance. It is worth noting that the policy only requires knowledge of the current network state, and do not require knowledge of the probabilities associated with future random events (Neely et al., 2005).

To facilitate modeling, we reiterate the problem here. Considering a roadway network with $N$ signalized intersections and time-varying traffic flows, each intersection $i \in \{1,2,\ldots,N\}$ is connected to a set of $\mathcal{L}(i)$ lanes (for inbound and outbound flows), where $\mathcal{L}(i) = \{l_1, l_2, \ldots, l_m\}$. The pair $(l_a, l_b)$ denotes the movement from lane $l_a$ through intersection $i$ to lane $l_b$, which is shown in **Fig. 1, where** $\{l_a, l_b\} \sim \mathcal{L}(i)$. Each intersection $i$ can be described by a tuple $(\mathcal{M}_i, \Phi_i, S_i)$, where $\mathcal{M}_i \subseteq \mathcal{L}^2(i)$ is the set of all possible movements through intersection $i$. Each flow movement through the intersection $i$ is defined by a pair $(l_a, l_b)$, where $(l_a, l_b) \in \mathcal{M}_i$. $\Phi_i$ is the set of all phases of signal at intersection $i$, $\phi_{ij} \in \Phi_i$ is phase $j$ controlling single or multiple movements at intersection $i$. We use the notation $(l_a, l_b) \in \phi_{ij}$ to indicate the phase $\phi_{ij}$ gives the right-of-way to the movement from lane $l_a$ through intersection $i$ to lane $l_b$. $s_{it}$ is the traffic state at intersection $i$ at time $t$. Clearly, a traffic state should be identified easily by observable attributes preferably through detectors around the intersection. In this study, a state at an intersection is represented by the current queues at the intersection. The following queue dynamics is used for each lane across this paper.

| | |
|---|---|
| $\Theta_i^l(t+1) = \max\left[\Theta_i^l(t) - \sum_b z_{ib}^{out}(t) + \sum_a z_{ai}^{in}(t) + A_i(t), 0\right]$ | (1) |

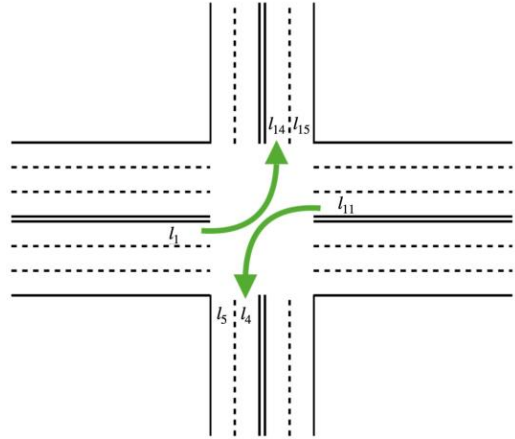

**Fig. 1.** illustrative flow movement for left-turn phase on a major arterial

Take intersection $i$ as an example. The time horizon $[0, T]$ is discretized into a set of timesteps $\{t_0, t_1, t_2, \ldots, t_k\}$. For each lane $l_a \in \mathcal{L}(i)$, $z_i(\phi_{ij}, l_a, l_b, s_t)$ denotes the flow rate that goes from $l_a$ to $l_b$ through intersection $i$ under state $s_{it}$ and phase $\phi_{ij}$. Note that $l_a$ is the lane through which traffic arrive at intersection $i$ while $l_b$ is an exit lane from intersection $i$, where $l_b$ is also an entry lane to a downstream intersection $j$. For simplicity, we abbreviate it as $z_{ab}$ in the following text. $z_{ab}$ has its value determined depending on the traffic situation. When there is a queue in lane $l_a$ at intersection $i$, $z_{ab}$ is equal to the saturation flow rate of the move through the intersection. Otherwise, it is equal to the arrival traffic flow rate from upstream(s).

Lyapunov control offers a comprehensive framework that can potentially unify various control policies in the transportation literature as explained in the subsequent subsections A and B below. Through different definitions of its function, Lyapunov control leads to different policies such that those maximizing intersection flow rate or minimizing the intersection total delay.

*Lyapunov function and flow-based control*

First consider DORAS, a policy focusing on switching the right-of-way to the phases with highest discharge efficiency. Details are available in (X. B. Wang et al., 2017). Here we present how Lyapunov control implies the DOARS policy.

We define $L(\Theta(t)) = \sum_{1 \le i \le N} \sum_{l \in \mathcal{M}_i} |\Theta_i^l(t)|$ as the Lyapunov function, representing a scale measure of the network congestion. For a given control policy and network at time $t$, the Lyapunov drift define as:

$$\Delta(\Theta(t)) = \mathrm{E}[|\Theta(t+1)| - |\Theta(t)|] \tag{2}$$

Because the queue $\Theta_i^l(t)$ is non-negative $\forall l$, we simplify the equation into the following form, $L(\Theta(t)) = \sum_{1 \le i \le N} \sum_{l \in \mathcal{M}_i} |\Theta_i^l(t)|$. Summing all intersection $i$ and all phase $l$, assuming queue is always exist at the intersection, we have

$$\Delta(\Theta(t)) = \sum_{1 \le i \le N} \sum_{l \in \mathcal{M}_i} [\Theta_i^l(t+1) - \Theta_i^l(t)]$$

$$= \sum_{1 \le i \le N} \sum_{l \in \mathcal{M}_i} \left[ \sum_a z_{ai}^{in}(t) + A_i(t) - \sum_b z_{ib}^{out}(t) \right] \tag{3}$$

Conduct the partial derivative on the right-hand side of equation with respect to the flow through the intersection $z_{ab}$, which represent the control as a phase. For a single intersection:

$$\frac{\partial \Delta(\Theta(t))}{\partial z_{ab}} = \frac{\partial}{\partial z_{ab}} \sum_{l \in \mathcal{M}_i} \left[ \sum_a z_{ai}^{in}(t) + A_i(t) - \sum_b z_{ib}^{out}(t) \right] = -\delta \tag{4}$$

The net change of drift, or the control-drift under control $\theta$ is $\{-\delta\} \cdot \{z_{ab}^\theta\}$. The derivative takes a value of -1 when the movement leads to the vehicle leaving network the network and $\delta$ (where $-1 \le \delta \le 0$) for movements within the network. Here $z_{ab}^\theta$ represents the flow of a moving through an intersection under control $\theta$, which can be considered akin to a signal plan. $z_{ab}^\theta$ equals to the saturation flow rate of the movement $(l_a, l_b)$ when the green signal is given to the approach with queued vehicles, and the arrival flow rate when the green signal is given to an approach without queued vehicles. $z_{ab}^\theta$ is zero when a red signal is given to the movement $(l_a, l_b)$. Note that the derivates of the RHS of Eq. (4) means that the gradient of the drift function decreases fastest along the approaches that discharge at the highest flow rates.

Viewing each intersection individually, Eq. (4) shows that a policy $\theta$ shall be adopted when it leads to the minimum drift. This is achieved when the phase that gives the maximum flow rate in the time interval gets the green signal. Essentially, the signal phase that discharges at the most total rate would decrease the drift function in the fastest rate, suggesting these approaches should be prioritized.

However, this is still a myopic view of control as it does not consider the causal effect over time and across connected intersections. Considering these factors, a more precise measure of the dispersion effect is needed to get $\delta$ estimation accurately. Under such constraints, the policy should be designed as maximizing the output flow of each intersection if each one is considered individually, which naturally leads to the concept of throughput maximization in DORAS, where it uses predicted flow within a cycle to estimate the discharge efficiency. The definition of $L\big(\Theta(t)\big) = \sum_{1 \leq i \leq N} \sum_{l \in \mathcal{M}_i} \big|\Theta_i^l(t)\big|$ might not fully capture network dynamics as it lacks consideration of interactions between intersection controls. It shown that Lyapunov function should be chosen wisely, although there are no excessive selection restrictions.

### Lyapunov function and pressure-based control

Next, we show how the Lyapunov control can lead to the popularly studied pressure-based control. We first define our particular Lyapunov function as in Eq. (5) before we show how it implies the back pressure method. More details about the back pressure method is seen in [18, 46].

$$L\big(\Theta(t)\big) = \sum_{i,l} \frac{\Theta_i^l(t)^2}{2} \tag{5}$$

The Lyapunov function has inherent connection to the intersection waiting time. In a deterministic queue for an approach $l$, one may roughly estimate the total vehicle waiting time to be $\frac{\Theta_i^l(t)^2}{4\lambda_l^\theta}$, where $\lambda_l^\theta$ is the discharge rate under policy $\theta$. To formally align the Lyapunov function with vehicle queueing time minimization, one might define the Lyapunov function as follow:

$$L\big(\Theta(t)\big) = \sum_{i,l} \frac{\Theta_i^l(t)^2}{8\lambda_l^\theta} \tag{6}$$

In the formulation presented in Eq. (6), a key aspect is the normalization of the squared queue term by the respective discharge rate $\lambda_l^\theta$ of each lane. For the ease of presentation, we choose to continue to use Eq. (6) for technical derivation while we keep in mind that $\lambda_l^\theta$ as a factor to differentiate the queues. Consequently, Eq (5) is proportional to the RHS of Eq. (6), both of which may serve for the purpose of vehicle waiting time minimization. In summary, the proposed Lyapunov function is inherently related to the intersection vehicle waiting time. This partly explains why our proposed control algorithm outperforms other algorithms. The corresponding drift function is derived as follows:

$$
\Delta\big(\Theta(t)\big) = E[L(\Theta(t+1)) - L(\Theta(t))] \tag{7}
$$

$$
= \sum_{1 \le i \le N} \sum_{l \in \mathcal{M}_i} \left[ \frac{1}{2}\left( \sum_a z_{ai}^{in}(t) + A_i(t) - \sum_b z_{ib}^{out}(t) \right)^2 + \Theta(t)\left( \sum_a z_{ai}^{in}(t) + A_i(t) - \sum_b z_{ib}^{out}(t) \right) \right]
$$

Consider the partial derivative on the Lyapunov drift with respect to a flow $z_{ab}$ through the intersection:

$$
\frac{\partial \Delta\big(\Theta(t)\big)}{\partial z_{ab}} = \sum_{1 \le i \le N} \sum_{l \in \mathcal{M}_i} \frac{\partial}{\partial z_{ab}} \left[ \frac{1}{2}\left( \sum_a z_{ai}^{in}(t) + A_i(t) - \sum_b z_{ib}^{out}(t) \right)^2 + \Theta(t)\left( \sum_a z_{ai}^{in}(t) + A_i(t) \right. \right. \tag{8}
$$

$$
\left. \left. - \sum_b z_{ib}^{out}(t) \right) \right]
$$

$$
= -\left( \sum_a z_{ai}^{in}(t) + A_i(t) - \sum_b z_{ib}^{out}(t) \right) - \Theta(t) + \left( \sum_a z_{aj}^{in}(t) + A_j(t) - \sum_b z_{jb}^{out}(t) \right) + \Theta(t)
$$

$$
= \Theta^{l_a}(t+1) - \Theta^{l_b}(t+1)
$$

Where $\partial z_{ab}$ indicates the flow rate through intersection. Then, we can have

$$
\partial \Delta^\theta \big(\Theta(t)\big) = \sum_{\forall (a,b)} \partial z_{ab}^\theta \left( \Theta^{l_a}(t+1) - \Theta^{l_b}(t+1) \right) \tag{9}
$$

Note here $\partial z_{ab}^\theta$ represents the incremental flow under policy $\theta$. It doesn't require a saturation flow and can be real-time discharge flow rate. Note that the policy $\theta$ may be interpreted as a sequence of a particular signal plan that comprises of the set of phases, sequence, and timing. For a particular phase $\phi_{ij} \in \Phi_i$, the resulting drift value associated with the intersection $i$ is expressed:

$$
E[\partial \Delta^\theta \big(\Theta(t)\big)|\phi_{ij}] = \sum_{(a,b) \in \phi_{ij}} \partial z_{ab}^\theta \left( \Theta^{l_a}(t+1) - \Theta^{l_b}(t+1) \right) \tag{10}
$$

In the set of phases $\Phi_i$ at each intersection, phases are mutually exclusive, activating one phase results in all other phases in $\Phi_i$ are red signal. Eq. (10) indicates that each phase $\phi_{ij} \in \Phi_i$ at the intersection state $s$ has its corresponding drift value. To minimize the drift function, the set of phases for the next timestep over the intersections should be chosen so that the drift value is minimize over all the intersections. There are two cases that govern the sequence of phases at intersections. One is a fixed sequence of phases, in which case, the control decision at any timestep would be whether it leads to a smaller drift value if the signal switches to the next phase. if so, the signal is switched to the next phase. Otherwise, the current phase continues. The second case does not have a fixed sequence of phases. All phases in $\Phi_i$ for intersection $i$ are examined in light of Eq. (10) in terms of the drift values. The phase with the smallest drift value is chosen to be the next phase to switch to.

The Lyapunov function utilized in this study adopts a myopic approach along the time dimension, focusing primarily on immediate outcomes rather than longer-term prediction beyond the next time epoch or adjacent intersections. This myopic perspective facilitates decoupling the network into a series of independent intersections, each controlled separately. Implementing the policy in a distributed manner has the advantage of simplifying the control process, where the expected differences in queues indicate the direction for minimizing the downward drift. This approach links the Lyapunov drift to the pressure series algorithm. The difference between queue is the pressure. Back pressure can be interpreted as the flow weighted pressure. Note that the estimating the queue length at the next step $t + 1$ may not always be straightforward because both the upstream and downstream queues may have flows fed from other entry/exit approaches. However, given the typically short duration of each time epoch, dramatic queue changes are unlikely and thus the estimate shall be feasible. We hereby propose to use the existing queue length to estimate such value and push the Lyapunov function to smaller value and stabilize the system at each time step.

### Network stability

The above sections assume the case that queue is deterministic and always exist, we will use drift method to show more general case, and the method fit for the stochastic situation. We consider the quadratic Lyapunov function $L\big(\Theta(t)\big) = \sum_{l \in \mathcal{M}_i} \Theta_n^l(t)^2$ and assume $\mathrm{E}[L(0)] < \infty$. There are constants $B > 0$, $\epsilon \geq 0$, such that the following drift condition holds for all timestep $t \in T$ and all possible $\Theta(t)$:

$$\Delta\big(\Theta(t)\big) \leq B - \epsilon \sum_{n,l} \Theta_n^l(t) \qquad (11)$$

if $\epsilon \geq 0$, all queues are mean rate stable, more strictly, if $\epsilon > 0$, all queues are strongly stable and all queues $\Theta_n^l(t)$ have

$$\lim_{t \to \infty} sup \frac{1}{t} \sum_{t=0}^{t-1} \sum_{1 \leq i \leq N} \sum_{l \in \mathcal{M}_i} \mathrm{E}[\Theta_n^l(t)] \leq \frac{B}{\epsilon} \qquad (12)$$

For any control policy, the Lyapunov drift at any timestep $t$ satisfies:

$$\Delta(\Theta(t)) \leq B - [\Phi(\Theta(t)) - \Gamma(\Theta(t))] \tag{13-a}$$

Where,

$$B = (z_{max}^{out})^2 + (z_{max}^{in} + A_{max})^2 \tag{13-b}$$

$$\Phi(\Theta(t)) = \sum_{i,l} \Theta_i^l(t) \mathrm{E}\left[\sum_b z_{ib}^{out}(t) - \sum_a z_{ai}^{in}(t) \,\middle|\, \Theta(t)\right] \tag{13-c}$$

$$\Gamma(\Theta(t)) = \sum_{i,l} \Theta_i^l(t) \mathrm{E}[\sum_c A_{ic}(t) | \Theta(t)] \tag{13-d}$$

$z_{max}^{out}$ and $z_{max}^{in}$ are the maximum input and output flow, respectively.

### Backpressure

Returning to the Lyapunov function as defined in Eq. (6), we introduce the concept of differential queue backlog $W_{ab}^{\phi} = \Theta_i^{l_a}(t) - \Theta_i^{l_b}(t)$ for a given movement $(l_a, l_b)$, assuming that the queue in $l_a$ is only discharged to the downstream queue in lane $l_b$. A general definition for $W_{ab}^{\phi}$ is as follows:

Under each phase $\phi_{ij}$ at intersection $i$, likely there are multiple concurrent traffic, each having a back pressure due to the queue differentials. A phase differential is therefore defined by including differentials for all the move at timestep $t$:

$$D_{\phi_{ij}}(t) = \sum_{(l_a, l_b) \sim \phi_{ij}} W_{ab} z_{ab}(\phi_{ij}, l_a, l_b, s_t) \tag{14}$$

Unlike max pressure policy using saturation flow, we use the Greenshields model to estimate the max flow rate through the intersection of each movement controlled by a phase at each timestep $t$:

$$z(t) = v_f d(t) - \left[\frac{v_f}{d_{jam}}\right] d^2(t) \tag{15}$$

The back pressure policy allocates the green time to the phase with max weighted pressure and thus use it to equalize differential queue backlog. Instead of making prediction on the arrivals, it uses flow on the lanes to incorporate the arrival information. The algorithm is similar in nature to back pressure for a communication network. In pervious works (Papageorgiou et al., 2003; Tassiulas & Ephremides, 1992; Wongpiromsarn et al., 2012), it haves been proven that back pressure routing leads to maximum network throughout.

In this study, we implement the DORAS-Q and Backpressure methods previously outlined and apply them to construct the reward function in Reinforcement Learning. All numerical experiments are conducted based on this framework.

*Reinforcement Learning*

In addressing the stochastic aspects and info that are not fully captured in the theoretical derivation, we employ Reinforcement Learning (RL). RL is adept at handling the variability and unpredictability inherent in traffic flows. The key components of our RL approach are defined as follows:

*State*: current phase $\phi_{ij}$, the total number of moving vehicles and stopped vehicles (speed < 0.1 m/s, or 0.22 mph) on each incoming lanes ($l_a$) and outgoing lanes ($l_b$).

*Action*: at each time $t$ each agent chooses a phase as its action $a_t$ from action set $A$, indicating the traffic signal should be set as current phase $\phi_{ij}$. Each action candidate $a_i$ is represented as a one-hot vector.

*Reward*: **for flow-based method,** the reward $R_t(t)$ for the intersections is the defined as the sum of negative absolute of all phase's flow.

$$R_t(t) = -\sum_{\phi_{ij} \in \Phi_i} \left| \sum_{(l_a, l_b) \sim \phi_{ij}} z_{ab}(\phi_{ij}, l_a, l_b, s_t) \right| \qquad (16)$$

*Reward*: **for pressure-based method,** the reward $R_t(t)$ for the intersections is the defined as the sum of negative absolute of all phase's backpressure.

$$R_t(t) = -\sum_{\phi_{ij} \in \Phi_i} \left| D_{\phi_{ij}}(t) \right| \qquad (17)$$

**Numerical Simulation result and discussion**

Numerical simulation has been conducted in two types of road network: a single corridor and a grid network. The reason for separate testing on a single corridor is that single corridors are often the major means of dealing with urban traffic. The settings are shown in **Fig. 2**. We have utilized SUMO 1.8 micro-simulator in conjunction with TraCI (Traffic Control Interface) 1.8 for modeling the case. SUMO (Simulation of Urban MObility) is an open-source, microscopic and continuous traffic simulation package designed to handle large traffic networks simulation with a large set of tools for scenario creation (Lopez et al., 2018). SUMO allows us to create a traffic simulation environment and track every vehicle. TraCI implements RL-based real-time signal control possible. The RL agent is built and trained in Pytorch 2.0 and Python 3.8.

*Environmental setting*

In the simulation, the link on minor arterial at each intersection consists of two through lanes. **Fig. 3** displays the phase timing plan for an arterial road network, which also outlines the available actions for the RL agent. Key timings include a 5-second amber interval (3 seconds yellow and 2 seconds all-red) to clear the intersection between phases, with green times ranging from a minimum of 5 seconds to a maximum of 50 seconds.

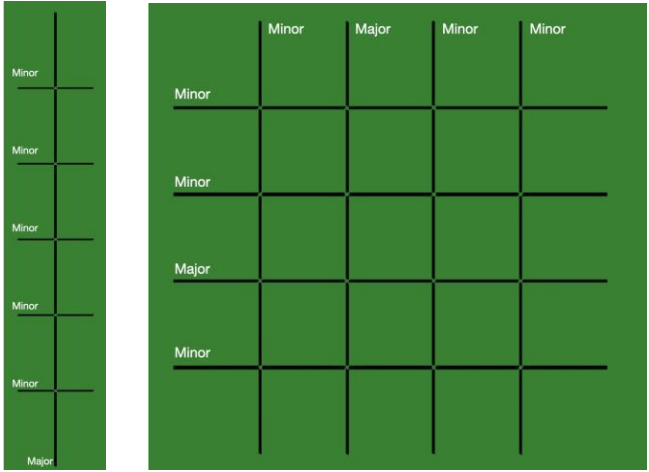

**Fig. 2.** Arterial and grid network simulation environment

**Fig. 3.** Phase plan for intersection of minor (left) and major (right)

TABLE I

DESIGNED TRAFFIC DEMAND IN THE SIMULATION

| Demand level | Major Road Traffic (vph) | Minor Road Traffic (vph) |
| --- | --- | --- |
| Low | 500 | 200 |
| Medium | 900 | 300 |
| High | 1300 | 400 |

*Baseline control algorithm*

We compare proposed model with two categorical methods: non-machine learning methods and RL-based model. In the RL-based model, we have used the waiting time, average queue length, and pressure as reward function as baselines of the RL-based model. Noted that all algorithms are fined-tuned, and RL-based model are uniformly trained by double DQN.

- Fixed-time (FT): This classical approach employs a fixed timing plan with green wave progression for coordinated traffic flow on arterial roads. In our study, fixed-timing plans and offsets are optimized using PASSER V.

- DORAS-Q: Designed for isolated intersection control, DORAS-Q can also be applied as a distributed control system across a network. Each intersection independently optimizes its control, gradually adapting the entire system. It utilizes existing queue lengths, short-term (typically 5 seconds), and average historical arrival rates for each phase to determine efficiency for phase switching.

- Max Pressure (MP): Max pressure defines the differences of queue between the current and the downstream intersection as pressure of the phase, and greedily selects the phase with the maximum pressure.

- Back Pressure (BP): The pressure is weighted dynamically by the traffic discharge rate and algorithm greedily chooses the phase with the maximum weighted pressure.

- RL-WaitingTime (RL-WT): Baseline of RL algorithms. The reward function is defined as waiting time and agent of RL aims to minimize the waiting time of all vehicles in the network.

- RL-Queue (RL-Q): Baseline of RL algorithms. The agent of RL aims to minimize the average queue length of each lane in the network.

- RL-MP: The RL algorithm aims to maximize the negative of absolute pressure.

- RL-MP: RL algorithm train to maximize the negative of absolute weighted pressure.

*Performance under homogenous traffic flow*

TABLE II

AVERAGE VEHICLE DELAY UNDER HOMOGENOUS PASSENGER CAR FLOW

|         | Low volume | | Medium volume | | High volume | |
|---------|-----------|---------|----------|---------|----------|---------|
|         | Arterial | Network | Arterial | Network | Arterial | Network |
| FT      | 30.93 | 97.97 | 38.06 | 139.87 | 89.82 | 191.16 |
| DORAS-Q | 23.25 | 72.77 | 36.64 | 84.82 | 76.64 | 148.92 |
| MP      | 19.88 | 54.17 | 31.59 | 73.26 | 71.62 | 157.61 |
| BP      | 18.99 | 44.75 | 26.92 | 69.56 | 56.86 | 137.82 |
| RL-WT   | 15.61 | **39.73** | 34.16 | 69.27 | 80.13 | 140.84 |
| RL-Q    | 15.38 | 40.67 | 26.47 | 60.88 | 66.47 | 140.87 |
| RL-MP   | 16.36 | 44.56 | 25.57 | 51.43 | 51.56 | 107.43 |
| RL-BP   | **13.46** | 42.95 | **22.64** | **47.75** | **46.07** | **92.08** |

*Performance under heterogeneous traffic flow*

TABLE III

AVERAGE VEHICLE DELAY UNDER 10% TRUCK HETEROGENOUS FLOW

|         | Low volume | | Medium volume | | High volume | |
|---------|-----------|---------|----------|---------|----------|---------|
|         | Arterial | Network | Arterial | Network | Arterial | Network |
| FT      | 36.96 | 99.26 | 52.72 | 153.79 | 132.38 | 226.39 |
| DORAS-Q | 28.87 | 74.82 | 49.76 | 96.09 | 130.47 | 184.25 |
| MP      | 20.96 | 50.20 | 41.50 | 85.99 | 125.14 | 184.25 |
| BP      | 19.87 | 46.69 | 37.14 | 73.74 | 113.65 | 148.83 |
| RL-WT   | **15.35** | 45.99 | 40.76 | 89.58 | 108.69 | 143.35 |
| RL-Q    | 15.45 | **43.56** | 32.21 | 88.17 | 106.03 | 147.22 |
| RL-MP   | 18.47 | 46.12 | 34.87 | 79.54 | 91.72 | 129.86 |

| RL-BP | 17.82 | 44.73 | **30.73** | **70.61** | **90.55** | **123.57** |

## TABLE IV
### Average vehicle delay under 25% truck heterogenous flow

|       | Low volume | | Medium volume | | High volume | |
|-------|-----------|---------|-----------|---------|-----------|---------|
|       | Arterial | Network | Arterial | Network | Arterial | Network |
| FT      | 42.13     | 100.62  | 84.94     | 181.88  | 148.03    | 258.91  |
| DORAS-Q | 31.14     | 75.38   | 57.52     | 128.95  | 118.36    | 227.39  |
| MP      | 23.18     | 51.76   | 41.92     | 100.83  | 122.27    | 203.20  |
| BP      | 21.95     | 50.27   | 36.58     | 90.15   | 119.61    | 191.65  |
| RL-WT   | 19.29     | **35.61** | 47.26   | 89.97   | 120.31    | 178.33  |
| RL-Q    | **17.78** | 36.23   | 35.39     | **87.11** | 119.78  | 176.39  |
| RL-MP   | 19.39     | 49.38   | 39.01     | 93.28   | 115.39    | 170.49  |
| RL-BP   | 18.50     | 46.63   | **34.58** | 87.26   | **113.48** | **167.27** |

## TABLE V
### Average vehicle delay under 40% truck heterogenous flow

|       | Low volume | | Medium volume | | High volume | |
|-------|-----------|---------|-----------|---------|-----------|---------|
|       | Arterial | Network | Arterial | Network | Arterial | Network |
| FT      | 69.04     | 106.25  | 129.64    | 192.63  | 267.68    | 274.39  |
| DORAS-Q | 34.02     | 76.68   | 120.33    | 146.14  | 225.39    | 259.32  |
| MP      | 27.93     | 65.68   | 127.96    | 110.89  | 239.37    | 259.45  |
| BP      | 23.25     | 63.38   | 117.83    | 98.46   | 216.21    | 240.86  |
| RL-WT   | 25.11     | 52.94   | 119.72    | 162.89  | 232.39    | 263.18  |
| RL-Q    | 23.74     | **51.14** | 114.71  | 165.79  | 232.79    | 247.28  |
| RL-MP   | 22.92     | 52.78   | 120.05    | 106.46  | 216.39    | 207.91  |
| RL-BP   | **22.18** | 51.49   | **111.05** | **95.87** | **208.72** | **187.45** |

## *Discussion*

The results are summarized from Tables II to V. RL-based algorithm generally exhibited strong performance in reducing total delay. It was particularly effective in medium and high-volume scenarios, surpassing other algorithms in efficiency. However, in low-volume situations, the clear advantage of the RL-based approach was less evident, although it still demonstrated satisfactory performance. Besides, traditional signal control methods perform much better with the addition of reinforcement learning algorithms because RL can leverage domain randomization to manage model uncertainty, offering more robust control responses, as well as enhanced flexibility and generality. (Degrave et al., 2022; Song et al., 2023)

In the fixed timing design, the green wave undoubtedly facilitates the vehicles' movement on the arterial and gird network at signalized intersections. However, this method lacks adaptability to dynamic traffic patterns and real-time intersection conditions, leading to inferior performance compared to more responsive systems.

DORAS-Q performs better than fixed-time control because each intersection could utilize the arrival steam information from nearby intersections. However, DORAS-Q, focusing on intersection efficiency and traffic prediction, encounters challenge due to potential inaccuracies in its forecasting. These inaccuracies, coupled with a lack of corrective mechanisms, often result in misaligned decisions with future traffic conditions. Additionally, vehicles usually arrive in the form of a platoon. The current phase drops rapidly to zero after clearing the queue, allowing the signal to switch to the next phase based on DORAS-Q. The tendency for rapid phase changes in DORAS-Q can disrupt its intended functionality, especially in grid network scenarios where intersections could otherwise benefit from shared traffic information. This issue is particularly pronounced in fully connected vehicle environments, emphasizing the importance of accurate data and real-time adaptability in traffic management – a core strength of RL-based approaches. On the other hand, the truck volume may decrease the algorithm's performance compared with the uniform passenger vehicle flow case. Using conversion factor to convert a truck to passenger vehicles in the queue length estimation may not work well in the algorithm. Although the algorithm is based on the queue length estimation, more truck characteristics may need to consider improving the algorithm.

The Max Pressure algorithm capitalizes on real-time intersection data, circumventing the limitations of predictive methods. However, its strategy of preferentially choosing the phase with maximum pressure tends to yield locally optimal solutions, a limitation that becomes more pronounced in complex road networks. Notably, in grid network scenarios, this approach often results in lower performance than DORAS-Q due to its inclination towards local rather than global optimization. Back pressure, by considering the potential flow through network nodes, refines queuing network control theory. Crucially, these pressure-based strategies require only the current state information, negating the need for predictive data and ensuring resilience against delays in data collection.

In simulations, Max Pressure algorithms showed sensitivity to traffic flow variations, especially under high-volume conditions. This variability under identical initial conditions highlighted the need for our approach of back pressure that adapts to current traffic flows. We also integrated back pressure within an RL framework, effectively compensating for the limitations of a standalone pressure series strategy. This integration facilitates iterative learning, steering the system toward an optimal solution that accounts for real-time traffic dynamics. This attribute enhances the system's responsiveness and efficiency, making RL-augmented pressure-based control a robust solution for dynamic traffic management. The only info required is the current state. A slight delay in the information collected will not affect the performance.

The inclusion of truck traffic in simulations typically leads to increased delays at intersections, with higher truck volumes correlating with longer delays. Two main reasons contribute to the increased delays caused by trucks: their larger size compared to passenger vehicles and their slower acceleration and deceleration rates. These factors lead to more heterogeneous traffic dynamics at intersections, challenging traditional traffic control methods. However, RL-based algorithms, particularly RL-BackPressure, show improved performance in high truck volume scenarios. While trucks introduce significant variability, affecting both performance and convergence time of RL algorithms, these algorithms are better suited to manage such challenges, particularly in scenarios with medium to high traffic volumes. This highlights the need for adaptable traffic control solutions capable of handling diverse vehicle behaviors efficiently.

Moreover, the stability of traffic movement at each intersection ensures the overall stability of the system. In road environments without turnarounds, the actions executed by RL agents avoid creating loops or blockages within the network. This approach effectively reduces imbalances across intersections, leading to more efficient utilization of green light time. Addressing a potential concern, one might wonder if equal long queues in every direction at an intersection, resulting in zero pressure, could worsen congestion. However, this scenario is typically transient and unstable. In such situations, when a phase turns green, the back pressure in that direction begins to decrease. Eventually, the RL agent will shift the green light to the next phase, seeking a higher reward, which in turn prompts the network towards reduced flow and pressure. Over a specified period , the RL agent is designed to maximize network throughput, thereby minimizing the overall travel time for all vehicles in the system. This capacity to adaptively manage traffic signals, even in complex scenarios, underscores the effectiveness of RL-based traffic control strategies.

**Conclusion**

Network intersection traffic control is a theoretically challenging and yet practically very significant problem. Currently, several measurements have been used for evaluating network signal performance, including but not limited to the total travel time of vehicles in the network, average vehicle delay and the number of stops, and average travel speed. Evaluating the performance of signal control usually needs more than one metric to validate each other. Thus, the fundamental problem of signal control is how to represent performance (i.e., minimum delay, minimum stops) through timely, observable, explicitly controllable variables. Addressing this, our study proposes a back pressure (weighted max pressure) method for network signal control, which is implied in the Lyapunov control function. This method considers the differential queues between queues before and after an intersection weighted by the traffic discharge rate.

Lyapunov control is a dynamic control framework whose objective is to minimize the drift function, which is usually not the objective function in the usual intersection signal control area. A significant advantage of Lyapunov control is that it maintains network stability under mild conditions, which is often most desired in the general network control area. We discover that Lyapunov control is a general, unifying umbrella. Through definition of particular Lyapunov functions, we specifically show that it implies the methods DORAS and back pressure.

In particular, we derive the back pressure (weighted max pressure) method for the first time through the Lyapunov function. In this regard, our proposed adoption of the reinforcement learning method demonstrates its great value in grasping the granular, often repetitive network effects such as the correlation between upstream and downstream traffic, platoon dispersion, etc as detailed in the beginning section of this paper. Our proposed method combines traffic theory and control theory and outperforms all other control methods that exist in the literature. The Lyapunov control is introduced in queuing control problems to reduce the impact of inaccurate prediction and increase the robustness to queuing estimation. The state of the RL algorithm is the total number of vehicles and stopped vehicles on each incoming and outgoing lane. The action is phase to allocate right-out-way. The reward for the intersection is the back pressure to reduce the imbalance across the network queues. The Double DQN was adopted as a function approximator with higher accuracy. The proposed signal control framework has solid theoretical support from control theory and is robust to distribution and minor errors. Moreover, the result shows that RL could improve the performance of traditional signal control algorithms. Training in randomized environments increases the robustness and adaptability of the agent, enhancing its effectiveness in managing disturbances and model uncertainties. RL-based algorithms have relatively better performance than fixed-time, DORAS-Q, Max Pressure, and Back Pressure in terms of total delay under varying volumes and in consideration of freight traffic. Among the RL-Based models, the proposed **RL-BackPresure** always has lower average vehicle delay in arterial and the grid network, especially under high volumes. In addition, the **RL-BackPresure** relies on the current state rather than predicting future states and it is embedded with physics information through effective reward design, so that the framework is more robust and accurate than current models.

Heterogeneous traffic, particularly that including significant truck traffic, is a situation worthy of great attention. Compared with traffic of uniform passenger flow, the truck traffic increases the overall delay at the intersection, and may deserve a special attention in signal control also in order to increase the economic efficiency. Freight traffic is more likely to disturb the downstream traffic potentially diminishing the performance of signal control algorithms. ]. However, RL-based algorithm remain robust and adeptly capture these disturbances.

There are a few limitations to this study. First, the performance of **RL-BackPressure** is not outstanding for light traffic loads in the case of both arterial and gird networks. It is worth investigating how to improve the algorithm when the small network effect exists. Second, a conversion factor may not fully represent trucks' characteristics. Considering detailed vehicle-specific characteristics and performance may facilitate a better comprehensive study and robust traffic signal control algorithm in future studies. It may be necessary to incorporate the Monte Carlo tree search or embed spatial information into the state, such as the state of adjacent intersection, into the algorithm to increase its robustness under various scenarios.

### Acknowledgment

This research is gratefully funded by the grant from the Federal Department of Transportation through the Freight Mobility Research Institute (FMRI).